\documentclass[showpacs,twocolumn,aps]{revtex4}
\usepackage{graphicx}
\usepackage{psfrag}
\usepackage{amsmath}

\usepackage{graphics}

\usepackage{epsfig}

\newcommand{\ee}{\hbox{e}}

\newcommand{\p}{\partial}

\newcommand{\kb}{k_{\hbox{\tiny B}}}
\begin{document}

\title{On the free energy within the mean-field approximation}
\author{Rafael Agra}
\affiliation{Laboratoire de Physique Th\'eorique
(UMR 8627 du CNRS), B\^atiment 210,
Universit\'e Paris-Sud, 91405 Orsay, France}
\author{Emmanuel Trizac}
\affiliation{Laboratoire de Physique Th\'eorique et Mod\`eles Statistiques
(UMR 8626 du CNRS), B\^atiment 100,
Universit\'e Paris-Sud, 91405 Orsay, France}
\author{Fr\'ed\'eric van Wijland}
\affiliation{Laboratoire de Physique Th\'eorique
(UMR 8627 du CNRS), B\^atiment 210,
Universit\'e Paris-Sud, 91405 Orsay, France}
\begin{abstract}
We compare two widespread formulations of the mean-field approximation, based on {\it minimizing}
an appropriately built mean-field free energy.
We use the example of the
antiferromagnetic Ising model
to show that one of these formulations does not guarantee the existence
of an underlying variational principle. This results in a severe failure
where
straightforward minimization of
the corresponding mean-field free energy leads to incorrect results.
\end{abstract}
\maketitle


In statistical physics the mean-field approximation is one of the most common and easy-to-use
frameworks. It is also one of the most powerful, and often the only
available one. It allows one to convert the study of a many-body problem of interacting degrees of
freedom into that of independent degrees of freedom.
There are several manners of performing a
mean-field approximation.
Our purpose is not to review them nor to discuss the corresponding well-documented
pitfalls \cite{chandler, diuetal, goldenfeld, lebellac,yeomans,parisi}. Among those
various mean-field versions, the one that is based on a rigorous variational principle
plays a special role. Our purpose in the present note is to confront this rigorous, albeit
cumbersome approach, with simpler and widely used formulations. We would like to analyze a hazardous
ambiguity in the concept of free energy in those apparently
more physical formulations which, to the best of our knowledge, has not been noted before. The reason
why these conceptually erroneous statements that can be found in many places in the literature
(\cite{diuetal}, \cite{tuckerman} ) have never been challenged is that pedagogical presentations
are usually confined to the ferromagnetic Ising model, which has a scalar order parameter,
and for which, somewhat luckily, the dangers we will point out remain hidden.

\begin{figure}[htb]
$$\input{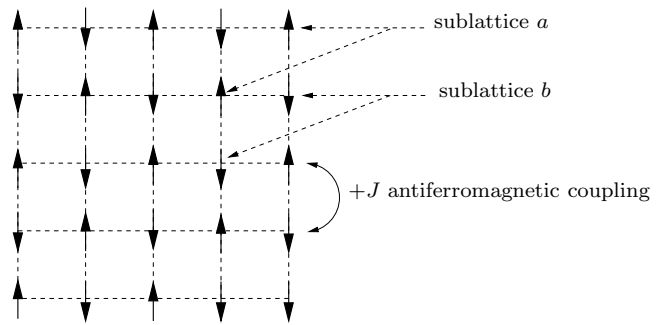}$$
\caption{A two dimensional anti-ferromagnet, for $T \ll T_N$.
The spins align almost perfectly in opposite directions on
two sub-lattices $a$ and $b$ of the original lattice.}
\label{sublatt}
\end{figure}

We have chosen to
illustrate our discussion with the  anti-ferromagnetic Ising model on a two-dimensional square lattice of $N$ sites, say
with periodic boundary conditions \cite{rque},
for our discussion would be pointless with a scalar order parameter.
The $N$ spins $s_i=\pm 1$, $i=1,...,N$ are
interacting according to the following Hamiltonian:
\begin{equation}
{\cal H}=+J\sum_{\langle i,j\rangle}s_i s_j
\end{equation}
where the sum $\sum_{\langle i,j\rangle}$ runs over the $2N$ distinct pairs of
nearest neighbour sites, and where the constant $J>0$ is the
anti-ferromagnetic coupling. The high
temperature phase of the system is paramagnetic
($m_i=\langle s_i\rangle$ vanishes).
As the temperature is decreased below the so called N\'eel temperature $T_N$
anti-ferromagnetic order sets in: The spins align in opposite directions on two
square sub-lattices, as shown in Figure \ref{sublatt}. The lattice sites are conveniently
divided into two subsets $a$ and $b$, as already depicted
on Figure \ref{sublatt}.\\

 We begin with a reminder of the variational formulation of the mean-field
approximation (route 1), and then present a more standard approach (route 2)
and the accompanying
difficulties in interpreting the related free energy.

\begin{figure}[h]
\begin{center}\epsfig{figure=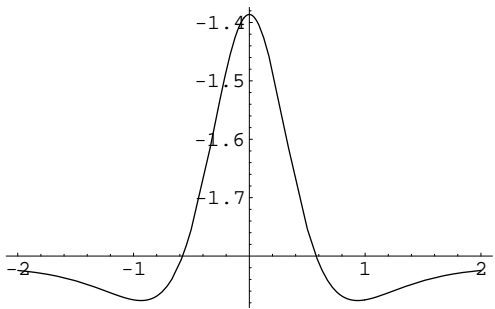,width=6.5cm}
\caption{Plot of the trial free energy
$\phi(m_a,m_b)$ at $ T=2.2 J/\kb<T_N$ as a function of $m_a$ along the
$m_a=-m_b$ direction. Note the
presence of  degenerate global minima at finite magnetization.\label{phi-coupe}}
\end{center}
\end{figure}

{\em Route 1: Variational procedure.} The idea~\cite{binneydowrickfischernewman,peierls,feynman} is to introduce a trial Hamiltonian ${\cal H}_{\text{mf}}$ depending on the original
degrees of freedom $\{s_i\}$ and on two parameters $m_a$ and $m_b$ the physical meaning of
which will become clear later. An intuitive choice for ${\cal H}_{\text{mf}}$ is
\begin{equation}\label{trialHmf}
{\cal H}_{\text{mf}}=+4J m_b\sum_{i\in a} s_i+4J m_a\sum_{j\in b}s_j
\end{equation}
Then one splits ${\cal H}$ into
\begin{equation}
{\cal H}={\cal H}_{\text{mf}}+({\cal H}-{\cal H}_{\text{mf}})
\end{equation}
so that the free energy $F$ of the system reads
\begin{equation}\label{bog}
F=F_{\text{mf}}-\kb T\ln \langle\ee^{-\beta({\cal H}-{\cal
H}_{\text{mf}})}\rangle_{\text{mf}}.
\end{equation}
Here, $\beta F_{\text{mf}} = -\log Z_{\text{mf}}$ where $Z_{\text{mf}}$
is the partition function associated to (\ref{trialHmf}),
$\beta = 1/(\kb T)$ is the inverse temperature with
$\kb$ the Boltzmann constant and
the angular brackets $\langle ...\rangle_{\text{mf}}$ denote an
average using the Gibbs measure
related to ${\cal H}_{\text{mf}}$, that is
with weight $\exp(-\beta {\cal H}_{\text{mf}})$.
The exponential being convex, one is led to the inequality
\begin{equation}
F\leq F_{\text{mf}}+\langle{\cal H}-{\cal H}_{\text{mf}}\rangle_{\text{mf}}.
\label{bogin}
\end{equation}
Equation~(\ref{bogin}) often appears under the name of Bogoliubov inequality
and may be used to find the best set of parameters $m_a$
and $m_b$ that render $\phi(m_a,m_b)=F_{\text{mf}}+\langle{\cal H}-{\cal
H}_{\text{mf}}\rangle_{\text{mf}}$ minimum, that is
as close as possible to the exact free energy $F$ \cite{rque2}.
It is important to note here that the best approximation
for $F$ is not $F_{\text{mf}}(m_a,m_b)$ but $\phi(m_a,m_b)$.
Using that $\langle s_{i\in a}\rangle_{\text{mf}}=-\tanh 4\beta J m_b$ and $\langle s_{j\in
b}\rangle_{\text{mf}}=-\tanh 4\beta J m_a$ we
arrive at
\begin{equation}\begin{split}
\beta \phi(m_a,m_b)=&\overbrace{\frac{N}{2}\Big[-\ln\left(4 \cosh 4\beta J m_a\cosh 4\beta J
m_b\right)}^{\beta F_{\text{mf}}}\\&+4\beta J\tanh 4\beta J m_a\tanh 4\beta J m_b\\
&+4\beta J\left(m_a \tanh 4\beta J m_a+m_b\tanh 4\beta J m_b\right)\Big]
\end{split}\end{equation}
Extremizing $\phi$ leads to the set of equations
\begin{eqnarray}\label{selfcon}
&&\frac{\p\phi}{\p m_a}=0,\;\frac{\p\phi}{\p m_b}=0\\
\Rightarrow && m_a=-\tanh 4\beta J m_b,\;m_b=-\tanh 4\beta J m_a
\end{eqnarray}
The latter system of equations has a unique solution $m_a=m_b=0$ at $\beta\leq \beta_N=\frac{1}{4J}$ ($T_N=4J/\kb$)
and possesses an additional set of two nonzero solutions for $\beta>\beta_N$ ($T_N$
is identified as the N\'eel temperature). In the high
temperature phase the paramagnetic solution $m_a=m_b=0$
becomes the global minimum of $\phi$, just as
the nonzero solution $m_a=-m_b\neq 0$ does in the low temperature phase (one can verify that the matrix of
the second derivatives of $\phi$ is positive definite at those extrema). Note also that
for
$\beta>\beta_N$ the paramagnetic state is a saddle point of $\phi$ with the unstable
direction along the $m_a=-m_b$ line (see Figure \ref{phi-coupe}). Right at the minimum, the expression of $\phi$ reads
\begin{equation}
\phi(m_a,m_b)= F_{\text{mf}}-2NJ m_a m_b\end{equation}
where $m_a$ and $m_b$ are the solutions to the system in (\ref{selfcon}).

At this stage we have simply postulated a trial Hamiltonian ${\cal H}_{\text{mf}}$ without
providing much of a physical motivation. It is {\it a posteriori} clear that ${\cal H}_{\text{mf}}$
describes a system of independent spins in an external magnetic field. For spin $s_i$ of
sub-lattice $a$ this magnetic field
is interpreted as the mean magnetization resulting from the four nearest neighbors on sub-lattice $b$, as
is confirmed by the fact that at the minimum of $\phi$ one can indeed verify that
\begin{equation}\label{otoko}
\langle s_{\in a}\rangle_{\text{mf}}=-\tanh 4\beta J m_b=m_a,\;\langle s_{\in
b}\rangle_{\text{mf}}=m_b
\end{equation}
In practice, however the variational procedure is not physically transparent and is mathematically
rather heavy. Furthermore it must be supplemented with a reasonable input of
physical intuition when postulating a trial Hamiltonian, lest the outcome of the calculation should be dull. Hence, for all these reasons, in spite of
$\phi(m_a,m_b)$ being a {\it bona fide} mean-field free energy,
it is rarely used in standard courses. The
purpose of the sequel is to present an alternative and widely used formulation of the mean-field
approximation~\cite{diuetal,tuckerman}, which at first glance appears more
satisfactory on physical grounds, but that conceals a number of hazards that we wish to
point out.

{\em Alternative formulation.}
Replacing in the original Hamiltonian $\cal H$ the spins $s_\ell$ with $m_{a\hbox{\tiny  ~or }b}+\delta s_\ell$
and neglecting terms quadratic in the $\delta s_\ell$'s, we obtain
our new mean-field Hamiltonian
${\cal H}'_{\hbox{\tiny mf}}$:
\begin{equation}\label{meanfieldH}
{\cal H}'_{\hbox{\tiny mf}}=-2NJm_a m_b+4Jm_b\sum_{i\in a}s_i+4Jm_a\sum_{j\in
b}s_j.
\end{equation}
In the present formulation, the mean-field approximation can be viewed as
neglecting correlations between nearest neighbour spin fluctuations.
The difference between the above ${\cal H}'_{\hbox{\tiny mf}}$ and the ${\cal
H}_{\hbox{\tiny mf}}$ that appears in (\ref{trialHmf}) lies in the additional constant
term $-2NJm_a m_b$ that features a temperature dependence through the magnetizations $m_a$
and $m_b$ that must carefully be kept track of. It is easily checked
that following the above variational procedure
route 1 with ${\cal H}'_{\hbox{\tiny mf}}$
instead of ${\cal H}_{\hbox{\tiny mf}}$ leads to the same results \cite{rque2}.
At this stage, another route can be followed, that differs from the
variational procedure. We decompose this second route into two steps.

{\em Route 2a: Self-consistency.}
From the mean-field Hamiltonian ${\cal H}'_{\hbox{\tiny mf}}$ it
is easy to deduce both the mean-field partition function
$Z'_{\hbox{\tiny mf}}$ and the
average  magnetization. We find
\begin{equation}\label{partitionfunction}
Z'_{\hbox{\tiny mf}}=2^N\ee^{2N\beta J m_a m_b}(\cosh(4\beta J m_a)\cosh(4\beta J m_b))^{N/2}
\end{equation}
and
\begin{eqnarray}\label{consistencya}
m_a=\langle s_{i\in a}\rangle=\frac{1}{Z'_{\hbox{\tiny mf}}}\sum_{\{s_\ell\}}s_i\ee^{-\beta {\cal H}'_{\hbox{\tiny mf}}}=-\tanh(4J\beta m_b)
\\
m_b=\langle s_{j\in
b}\rangle=-\tanh(4J\beta m_a)
\label{consistencyb}
\end{eqnarray}
This system of equations is exactly the one found in (\ref{otoko}). The self-consistency
equations (\ref{consistencya}) and (\ref{consistencyb})
have only the paramagnetic solution  when $T\geq T_N$, while a nonzero solution
continuously develops as $T$ is lowered below $T_N$. It is then argued that
below $T_N$, which is identified with the N\'eel temperature,
the $m_a=m_b=0$ solution is {\it unstable} while the solution $m_a=-m_b\neq 0$
becomes {\it stable} and is the physically relevant one. Either more precise
discussions about stability issues are discarded
or one finds in standard textbooks the following
assertion to justify the choice of the nonzero solution below $T_N$:
It becomes stable below the
N\'eel temperature (this is true), as can be checked by studying
the minima of the free energy. This is this last sentence that we would now like to discuss.

{\em Route 2b: Free energy landscape.}
From the expression of the partition function given in (\ref{partitionfunction}) one can easily
deduce an expression for the free energy $\phi'(m_a,m_b)$ as a function of the magnetizations on the two sub-lattices:
\begin{equation}\label{F}\begin{split}
\phi'(m_a,m_b)
=&-\kb T \log Z'_{\text{mf}}
=N\Big(-2Jm_am_b\\&-\frac{1}{2\beta}\ln\left[4\cosh(4\beta J m_a)\cosh(4\beta J
m_b)\right]\Big)
\end{split}\end{equation}
We now express that we search for the states that
{\it minimize} the free energy $\phi'$
\begin{equation}\label{derivativeF}
\frac{\p \phi'}{\p m_a}=\frac{\p \phi'}{\p m_b}=0
\end{equation}
Within the framework of the simpler ferromagnetic case, this is precisely the wording adopted {\it
e.g.} in \cite{diuetal}. It is then usually commented that the equations
(\ref{derivativeF}) are equivalent to those obtained
by resorting directly to the self-consistency
conditions. A plot of the free energy landscape as a function of the order
parameter usually follows. And indeed for $T <T_N$ it may be seen that the paramagnetic state
becomes a global maximum of $\phi'$ considered as a function of
independent variables $(m_a,m_b)$.

\begin{figure}[h]
\begin{center}\epsfig{figure=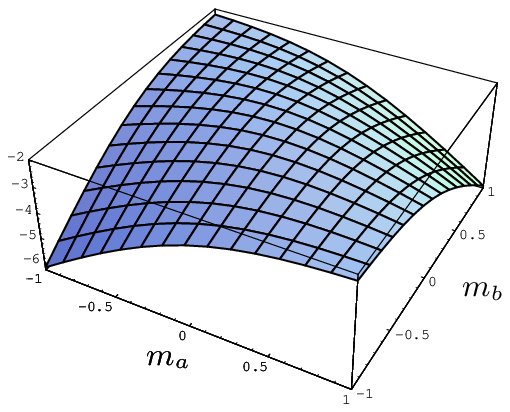,width=9.5cm}
\caption{Plot of the dimensionless free energy
$\phi'_{\hbox{\tiny mf}}/(JN)$ as a function of $m_a$ and $m_b$ in
$[-1,1]^2$, below the N\'eel temperature at $T=3\,T_N/4$.
\label{plotfreeenergy}}
\end{center}
\end{figure}

\begin{figure}[htb]
   \vspace{1.cm}
$$\epsfig{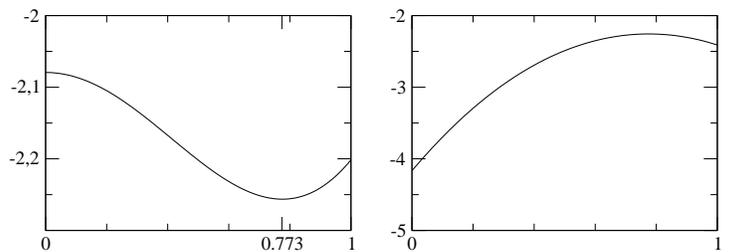}$$
\caption{Plot of the dimensionless free energy
$\phi'_{\hbox{\tiny mf}}/(JN)$ as a function of $m_a$
along the $m_a=-m_b$
line (left) and along the perpendicular direction crossing the saddle point at $m_a=-m_b\simeq
0.77$ (right). Same parameters as in Fig. \ref{plotfreeenergy}
\label{coupe}.}
\end{figure}

However, below the N\'eel temperature, the nontrivial state $(m_a,m_b)$
deduced from
(\ref{consistencya}) and (\ref{consistencyb}) is simply neither a local nor a global minimum of the free energy
$\phi'(m_a,m_b)$:\\
\noindent (i) there exist other states,
at the boundaries of
the magnetization domain, that have a lower free energy;\\
\noindent (ii) $(m_a,m_b)$ as given by the
nonzero solution of (\ref{consistencya}) and (\ref{consistencyb})
does not even correspond to a local minimum.\\
\noindent It is instructive to examine the shape of the free energy landscape as a function of the order parameter components $(m_a,m_b)$, as plotted in
Figure \ref{plotfreeenergy}.
We find that the state that globally minimizes the free energy $\phi'$ is the fully ordered {\it ferromagnetic}
$m_a=m_b=1$ state (or equivalently $m_a=m_b=-1$), whatever $0\leq T <T_N$.
Furthermore, the anti-ferromagnetic state corresponds to a saddle point of the free
energy landscape. This is best appreciated on Figure \ref{coupe}. With the chosen parameters the
correct anti-ferromagnetic state has $m_a\simeq 0.77$.
We have clearly come across an unexpected hazard of the mean-field
approximation. Finally, note that by artificially dividing a regular ferromagnetic Ising model on a square lattice
into two sub-lattices with independent average magnetizations, one would come across the same kind of problems for the
mean-field free energy (the ferromagnetic state would become a saddle-point, the
global minimum would correspond to the fully anti-ferromagnetic state, {\it
etc.}).

{\em Discussion.} We now come back to the variational formulation and we wish to underline
that when one evaluates $\phi$ at its global minimum one finds that
\begin{equation}
\phi(m_a,m_b)=\phi'(m_a,m_b)
\end{equation}
where $m_a$ and $m_b$ are the functions of temperature solution to the system
(\ref{consistencya},\ref{consistencyb}) or (\ref{otoko}). This tells us that
it is perfectly legitimate to follow route 2a
using the numerical value of $\phi'$ at the
values given by (\ref{consistencya},\ref{consistencyb}) for
finding the physical solution to the problem.
In other words, route 2a with the further computation of $\phi$ for
the self-consistent magnetizations is correct.
However it should be prohibited to freely vary the magnetizations
$m_a$ and
$m_b$ as in route 2b
to study the free-energy $\phi'$ landscape and to rely on the latter landscape to discuss stability issues.
There is indeed no variational principle underlying the derivation
of $\phi'$. This is at variance with the safe route 1 relying
of $\phi$. This also means that route 2 can never be used to discuss
metastability issues, even restricting to self-consistent magnetizations.

The inconsistency we have brought forth for route 2b
appears only when the mean-field order parameter
is not a scalar (as is the case in the ferromagnetic model): Ordering phenomena 
on more complex
substructures of the original lattice would inevitably lead to similar results.
The key point in our second formulation of the mean-field approximation
(route 2) is that
by their very definition, $m_a$ and $m_b$ are the average magnetizations:
They cannot be
considered as freely varying variables. They are functions of the temperature determined
by the self-consistency equations. It is curious to note that the physical
solution ($m_a,m_b$) to the problem (which is by definition a minimum
of $\phi$)
corresponds to a saddle point
of the function $\phi'$.

Therefore we would like to conclude by warning that there is in principle no physical
meaning to the mean-field free energy $\phi'$ seen as a function of a freely varying
order parameter. Comparing mean-field free energies is meaningful only for
solutions of the self-consistency equations. We have shown this below the N\'eel
temperature, but similar problems arise in the high-temperature limit. Indeed, expanding
the free energy Eq.~(\ref{F}) in the vicinity of $\beta=0$ yields $\phi'_{\hbox{\tiny
mf}}\simeq -2 J N m_a m_b$ from
which one could be tempted to conclude erroneously that the stable state is fully
ferromagnetic when it is of course paramagnetic!

\begin{thebibliography}{99}
\bibitem{rque2}
We note that the Bogoliubov bound [right hand side of Eq. (\ref{bogin})]
is the same for all trial Hamiltonians of the form ${\cal H}_{\text{mf}}+h(m_a,m_b)$
where $h(m_a,m_b)$ is an arbitrary function.
\end{thebibliography}
\end{document}